\documentclass[aps,twocolumn,showpacs,groupedaddress]{revtex4}
\usepackage{graphicx}
\newcommand{\Tr}{\text{Tr}}

\begin{document}

\title{Revisiting the Excitation Energy Transfer  in the Fenna-Matthews-Olson Complex}
\author{X. X. Yi$^{1,2}$, X. Y. Zhang$^{1}$,  and C. H. Oh$^2$}
\affiliation{$^1$School of Physics and Optoelectronic Technology\\
Dalian University of Technology, Dalian 116024 China\\
$^2$Centre for Quantum Technologies and Department of Physics,
National University of Singapore, 117543, Singapore }

\date{\today}

\begin{abstract}
It is believed that  the quantum coherence itself cannot explain the
very high excitation energy transfer (EET) efficiency  in the
Fenna-Matthews-Olson (FMO) complex. In this paper, we show that this
is not the case if the inter-site couplings take complex values.  By
phenomenologically introducing   phases into the inter-site
couplings, we  obtain the EET efficiency as high as 0.8972 in
contrast to 0.6781 with real inter-site couplings. Dependence of the
excitation energy transfer efficiency on the initial states is
elaborated. Effects of fluctuations in the site energies and
inter-site couplings are also examined.
\end{abstract}

 \maketitle

\section{introduction}

In 1962, John Olson isolated a water-soluble bacteriochlorophyll
({\it BChl a}) protein from green sulfur bacteria \cite{olson63}. In
1975, Roger Fenna and Brian Matthews resolved the X-ray structure of
this protein (Fenna-Matthews-Olson  (FMO)) from {\it
Prosthecochloris aestuarii} and found that the protein consists of
three identical subunits related by $C_3$ symmetry, each containing
seven {\it BChl a} pigments \cite{fenna75}.  In photosynthetic
membranes of green sulfur bacteria, this protein channels the
excitations from the chlorosomes to the reaction center. Since it
was the first photosynthetic antenna complex of which the X-ray
structure became available, it triggered a wide variety of studies
of spectroscopic and theoretical nature, and it therefore has become
one of the most widely studied and well-characterized
pigment-protein complexes.

The excitation-energy transfer can be elucidated with 2D
echo-spectroscopy using three laser pulses which hit the sample
within several femtoseconds time spacings. The experimentally
measured 2D echo-spectra of the complex show wave-like energy
transfer with oscillation periods roughly consistent with the
eigenenergy spacings  of the excited system\cite{engel07}. This has
brought a long-standing question again into the scientific focus
that whether nontrivial quantum coherence effects exist in natural
biological systems under physiological conditions
\cite{lee07,adolphs06}.

Many studies have attempted to unravel the precise role of quantum
coherence in the EET of light-harvesting complexes
\cite{mohseni08,plenio08,caruso09,chin10,olaya08,ishizaki09,yang10,hoyer09,sarovar11,shi11}.
The interplay of coherent dynamics, which leads to a delocalization
of an initial excitation arriving at the FMO complex from the
antenna, and the coupling to a vibronic environment with slow and
fast fluctuations, has led to studies of environmentally assisted
transport in the FMO. An interesting finding is that the
environmental decoherence and noise  play  a crucial role in the
excitation energy transfer in the FMO
\cite{mohseni08,plenio08,caruso09,chin10,caruso101,shabani11,ghosh11,yi01}.

In these studies, the FMO complex is treated using the  so-called
Frenkel exciton Hamiltonian,
\begin{equation}
H=\sum_{j=1}^7 E_j|j\rangle\langle
j|+\sum_{i>j=1}^7J_{ij}(|i\rangle\langle j|+h.c.),
\end{equation}
where $|j\rangle$ represents the state with only the $j$-th site
being excited and all other sites in their electronic ground state.
$E_j$ is the on-site energy of site $j$, and $J_{ij}$ denotes the
inter-site  coupling between sites $i$ and $j$. The inter-site
couplings $J_{ij}$ given by $J_{ij}=\int
d\vec{r}\Phi_i(\vec{r})\rho_j(\vec{r})$ represent  the Coulomb
couplings between the transition densities of the BChls
\cite{adolphs08}, where $\Phi_i(\vec{r})$ is the electrostatic
potential  of the transition density of {\it BChl i} and
$\rho_j(\vec{r})$ is the transition density of {\it BChl j}.  It is
easy to find that this calculation can only give real inter-site
couplings $J_{ij}.$ In fact, the couplings, which has been used in
previous studies of 2D echo-spectra and facilitate comparisons
between different approaches, are all real.

We should emphasize that the inter-site couplings originate from
dipole-dipole couplings between different chromophores. They are in
general complex from the aspect of quantum mechanics, regardless  it
is difficult to calculate.   In this paper, we will shed light on
the effects of  complex inter-site couplings by addressing  the
following questions. First, we present a theoretical framework which
elucidates how complex couplings  could assist the excitation
transfer. We optimize the relative phases in the couplings  for
maximal transfer efficiency and study the robustness of excitation
transfer against the fluctuation of the on-site energy and
inter-site couplings. Second, we present a study on the coherence in
the initially excited states by showing the dependence of the
transfer efficiency on the initial states.

This paper is organized as follows. In Sec. {\rm II}, we introduce
the model  to describe the FMO complex for the dynamics of the
exciton transport. In Sec. {\rm III}, we optimize the phases
phenomenologically introduced  in the couplings  for maximal
excitation energy transfer efficiency and explore how the efficiency
depends on the initial state of the FMO. The fluctuations in the
local on-site energy and in the inter-site couplings affect the
transfer efficiency, this effect is studied in Sec. {\rm IV}.
Finally, we conclude  in Sec. {\rm V}.

\section{model}
Since the FMO complex is arranged as a trimer with three different
subunits interacting weakly with each other, we  restrict our study
to a single subunit which contains eight bacteriochlorophyll
molecules (BChls)\cite{tronrud09, busch11}. The presence of the 8th
BChl chromophore has just been suggested by the recent
crystallographic data\cite{tronrud09}, and the recent experimental
data and theoretical studies indicated that the 8th BChl is the
closest site to the baseplate and should be the point at which
energy flows into the FMO complex\cite{busch11,ritschel11}. However,
the eighth BChl is loosely bound, it usually detaches from the
others when the system is isolated from its environment to perform
experiments. Thus we do not take the eighth into account and the EET
will be explored by  using the seven-site model\cite{tronrud09}.

We describe the FMO complex by a generic one-dimensional Frenkel
exciton model, consisting of a regular chain of 7 optically active
sites, which are modeled as two-level systems with parallel
transition dipoles. The corresponding Hamiltonian reads,
\begin{equation}
H=\sum_{j=1}^7 E_j|j\rangle\langle
j|+\sum_{i>j=1}^7(J_{ij}|i\rangle\langle j|+J_{ij}^*|j\rangle\langle
i|),
\end{equation}
where $|j\rangle$ represents the state where only the $j$-th site is
excited and all other sites are in their electronic ground state.
$E_j$ is the on-site energy of site $j$, and $J_{ij}$ denotes the
excitonic coupling between sites $i$ and $j$, $J_{ij}=J_{ji}^*$. The
on-site  and the inter-site energies from Ref.\cite{adolphs06} (in
units of $\mbox{cm}^{-1}$) will be adapted to study the quantum
dynamics of the FMO. As afromentioned, a phase $\phi_{ij}$ is added
to the inter-site coupling $J_{ij}$ stronger than 15 $cm^{-1}$,
\begin{widetext}
\begin{equation}
H \!=\!\! \left(\!\!\begin{array}{ccccccc}
\mathbf{215}   & \!\mathbf{-104.1e^{i\phi_{12}}} & 5.1  & -4.3  &   4.7 & \mathbf{-15.1e^{i\phi_{16}}} &  -7.8 \\
\!\mathbf{-104.1e^{-i\phi_{12}}} &  \mathbf{220.0} &\mathbf{ 32.6e^{i\phi_{23}}} & 7.1   &   5.4 &   8.3 &   0.8 \\
5.1 &  \mathbf{ 32.6e^{-i\phi_{23}} }&  0.0 & \mathbf{-46.8e^{i\phi_{34}}} &   1.0 &  -8.1 &   5.1 \\
-4.3 &    7.1 &\!\mathbf{-46.8e^{-i\phi_{34}}} & \mathbf{125.0} &\!
\mathbf{-70.7e^{i\phi_{45}}} &\! -14.7 &
\mathbf{ -61.5e^{i\phi_{47}}}\\
4.7 &    5.4 &  1.0 & \!\mathbf{-70.7e^{-i\phi_{45}}} & \mathbf{450.0} & \mathbf{ 89.7e^{i\phi_{56}}} &  -2.5 \\
\mathbf{-15.1e^{-i\phi_{16}}} &    8.3 & -8.1 & -14.7 &  \mathbf{89.7e^{-i\phi_{56}}} & \mathbf{330.0} & \mathbf{ 32.7e^{i\phi_{67}}} \\
-7.8 &    0.8 &  5.1 & \mathbf{-61.5e^{-i\phi_{47}}} &  -2.5 &
\mathbf{32.7e^{-i\phi_{67}}} & \mathbf{280.0}
\end{array}\!\!
\right). \label{ha}
\end{equation}
\end{widetext}
Here the zero energy has been shifted by 12230 $\mbox{cm}^{-1}$ for
all sites, corresponding to a wavelength of $\sim 800 \mbox{nm}$. We
note that in units of  $\hbar=1$, we have 1 ps$^{-1}$=5.3 cm$^{-1}$.
Then by dividing $|J_{ij}|$ and $E_{j}$ by 5.3, all elements of the
Hamiltonian  are rescaled by units of ps$^{-1}$. We can find from
the Hamiltonian $H$ that in the Fenna-Matthew-Olson complex (FMO),
there are two dominating exciton energy transfer (EET) pathways:
$1\rightarrow 2\rightarrow 3$ and $6\rightarrow (5,7)\rightarrow 4
\rightarrow 3$. Although the nearest neighbor terms dominate the
site to site coupling, significant hopping matrix elements exist
between more distant sites. This indicates that coherent transport
itself may not explain why the excitation energy transfer is so
efficient.

We phenomenologically introduce a decoherence scheme to describe the
excitation population transfer from site 3 to the reaction center
(site 8). The  master equation to describe the dynamics of the FMO
complex  is then given by,
\begin{eqnarray}
\frac{d\rho}{dt} = -i[H,\rho]+ {\cal L}_{38}(\rho)\;,
\label{masterE}
\end{eqnarray}
where $\mathcal{L}_{38}(\rho)=\Gamma \left[P_{83}\rho(t)
P_{38}-\frac{1}{2}P_3\rho(t)-\frac{1}{2}\rho(t)P_3\right]$ with
$P_{38}=|3\rangle\langle 8|=P_{83}^{\dagger}$ characterizes the
excitation trapping at site 8 via site 3.

We shall use the population $P_8$ at time $T$ in the reaction center
given by $P_8(T)=\Tr(|8\rangle\langle 8|\rho(T))$ to quantify the
excitation transfer efficiency. Clearly, the Liouvillian
$L_{38}(\rho)$ plays an essential role in the excitation transfer.
With this term, we will show in the next section that the complex
inter-site coupling can enhance the excitation energy transfer.

\section{simulation results}
It has been shown that the Hamiltonian with real inter-site
couplings can not give high excitation energy transfer efficiency.
Indeed, the previous  numerical simulations shown that $\Gamma$ can
be optimized to 87.14 with all $\phi_{ij}=0$ to obtain the highest
excitation transfer efficiency $p_8=0.6781$ in this case
\cite{yi12}.

\begin{figure}
\includegraphics*[width=0.8\columnwidth,
height=0.5\columnwidth]{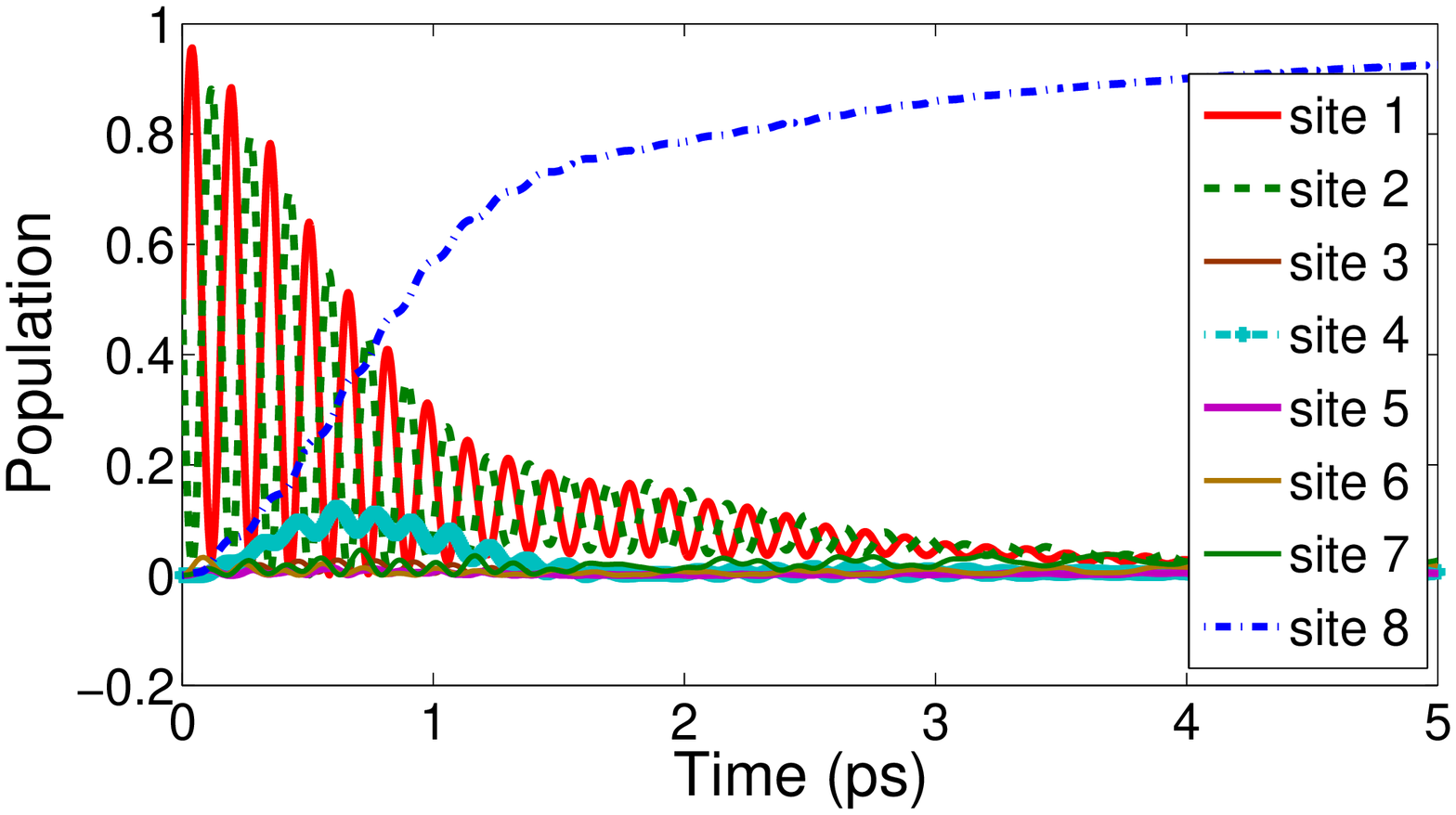}
\includegraphics*[width=0.8\columnwidth,
height=0.5\columnwidth]{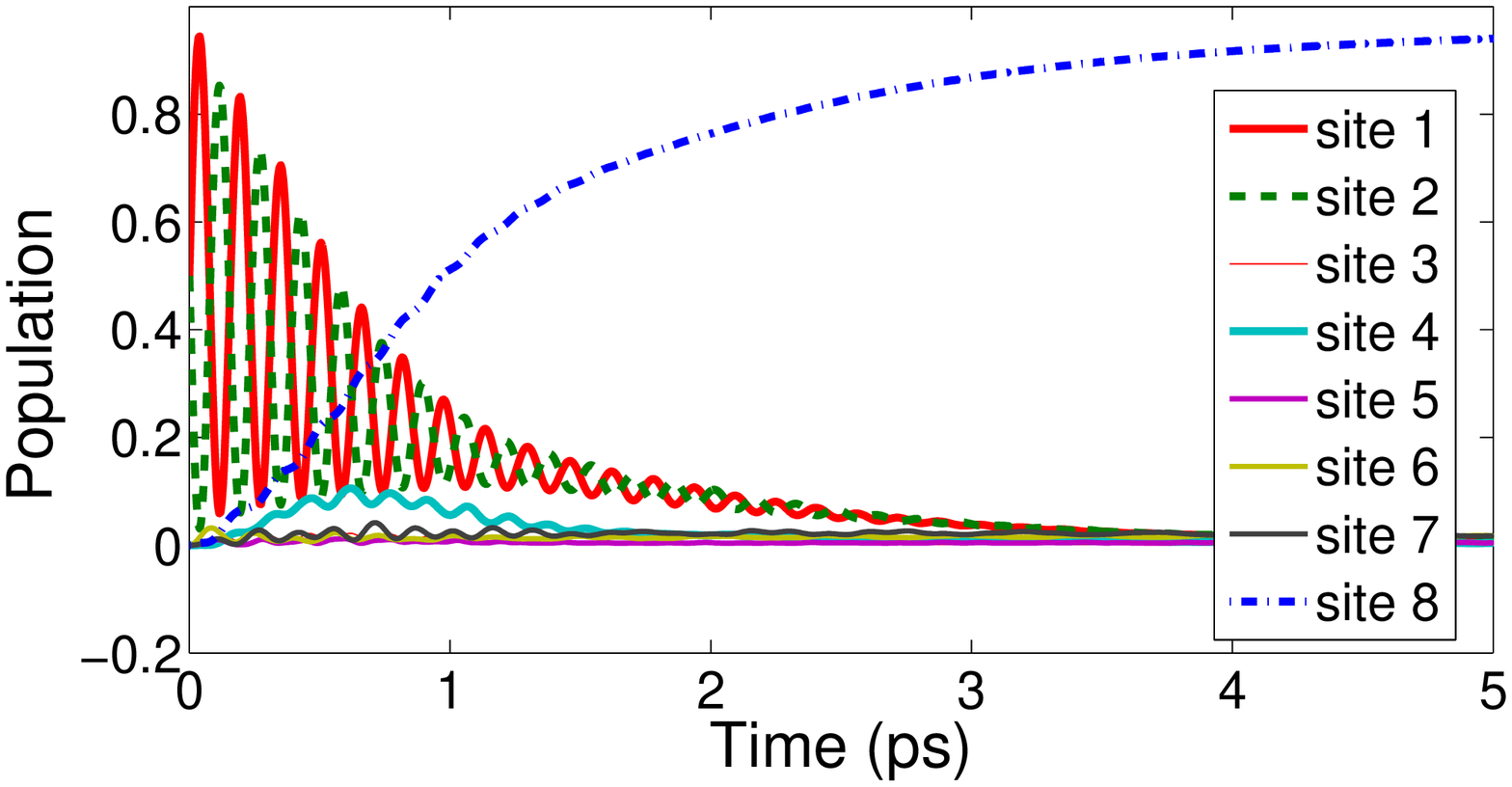} \caption{Population on each
site as a function of time. Site 8 represents the reaction center.
The phases are optimized for the EET efficiency. The exciton is
assumed initially on the site 1. The upper panel shows the
population without decoherence, while the bottom panel shows that
with $\gamma_2=2.4$ and $\gamma_{j\neq 2}=0,$ see
Eq.(\ref{masterE1}). $\Gamma=44$ is chosen throughout this paper.}
\label{fig1}
\end{figure}

With complex inter-site couplings, the EET efficiency can reach
almost $90\%$ by optimizing the phases added to $J_{ij}$. In units
of $\pi$, these phases are  $\phi_{12}=1.2566, \phi_{16}=3.3510,
\phi_{23}=1.8850, \phi_{34}=1.8850, \phi_{45}=1.8850,
\phi_{47}=0.8378, \phi_{56}=3.1416, \phi_{67}=0.3142.$ With these
phases, the population on each site as a function of time is plotted
in Fig. \ref{fig1}. Two observations can be made from Fig.
\ref{fig1}: (1) The population shows oscillatory behavior in the
whole energy transfer process, this is different from the situation
where decoherence dominates the mechanism of EET \cite{yi12},  (2)
the excitation on site 1 and 2 dominates the population, while the
population at site 3 is almost zero. The first observation can be
understood as the quantum coherence lasts the whole EET, and the
second observation suggests that the site 1 and site 2 play
important role in the excitation energy transfer. The excitation on
site 3 is trapped and transferred to the reaction center almost
immediately as to obtain a high transfer efficiency, as Fig.
\ref{fig1} shows.

\begin{figure}
\includegraphics*[width=0.8\columnwidth,
height=0.5\columnwidth]{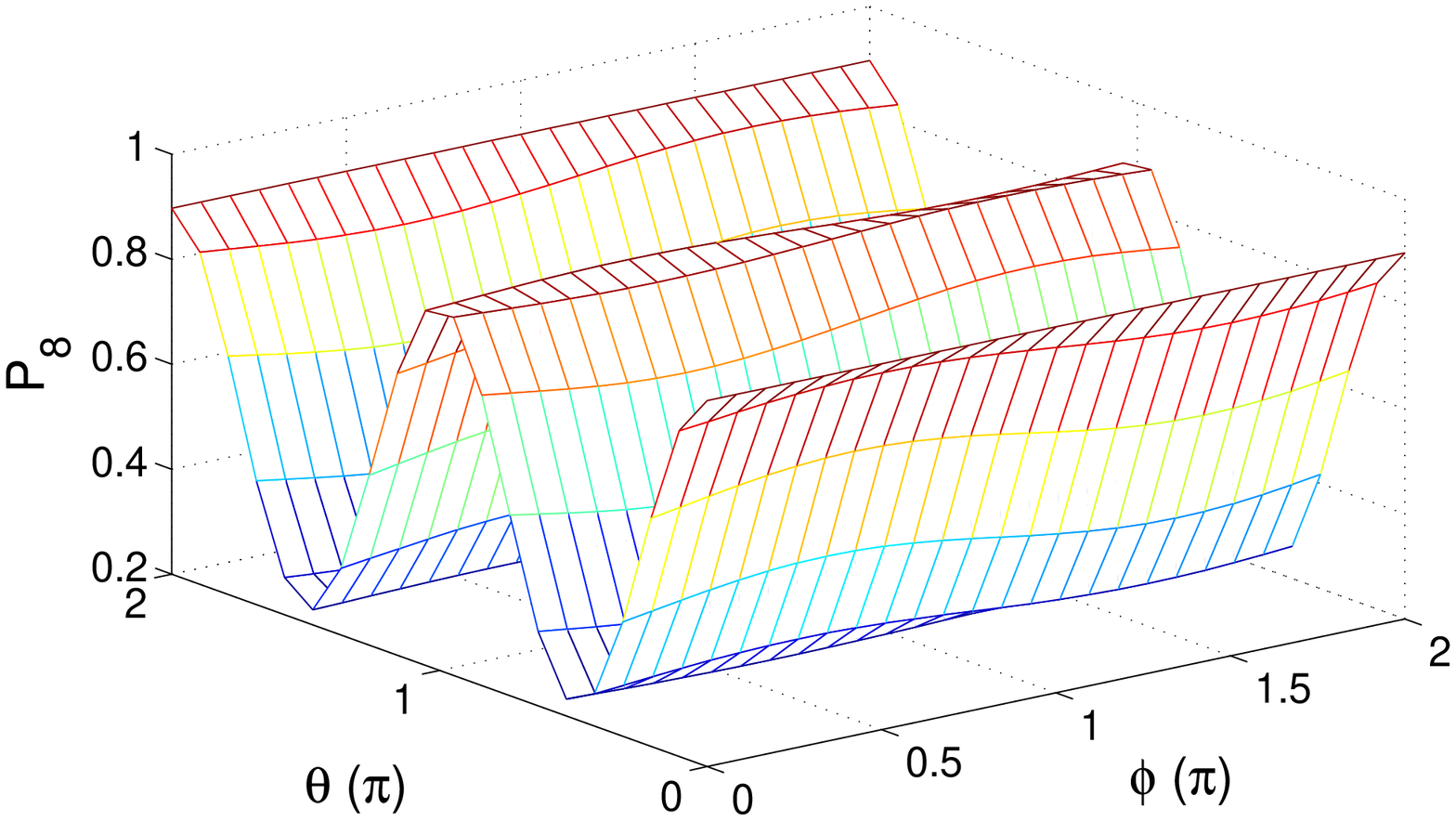}
\includegraphics*[width=0.8\columnwidth,
height=0.5\columnwidth]{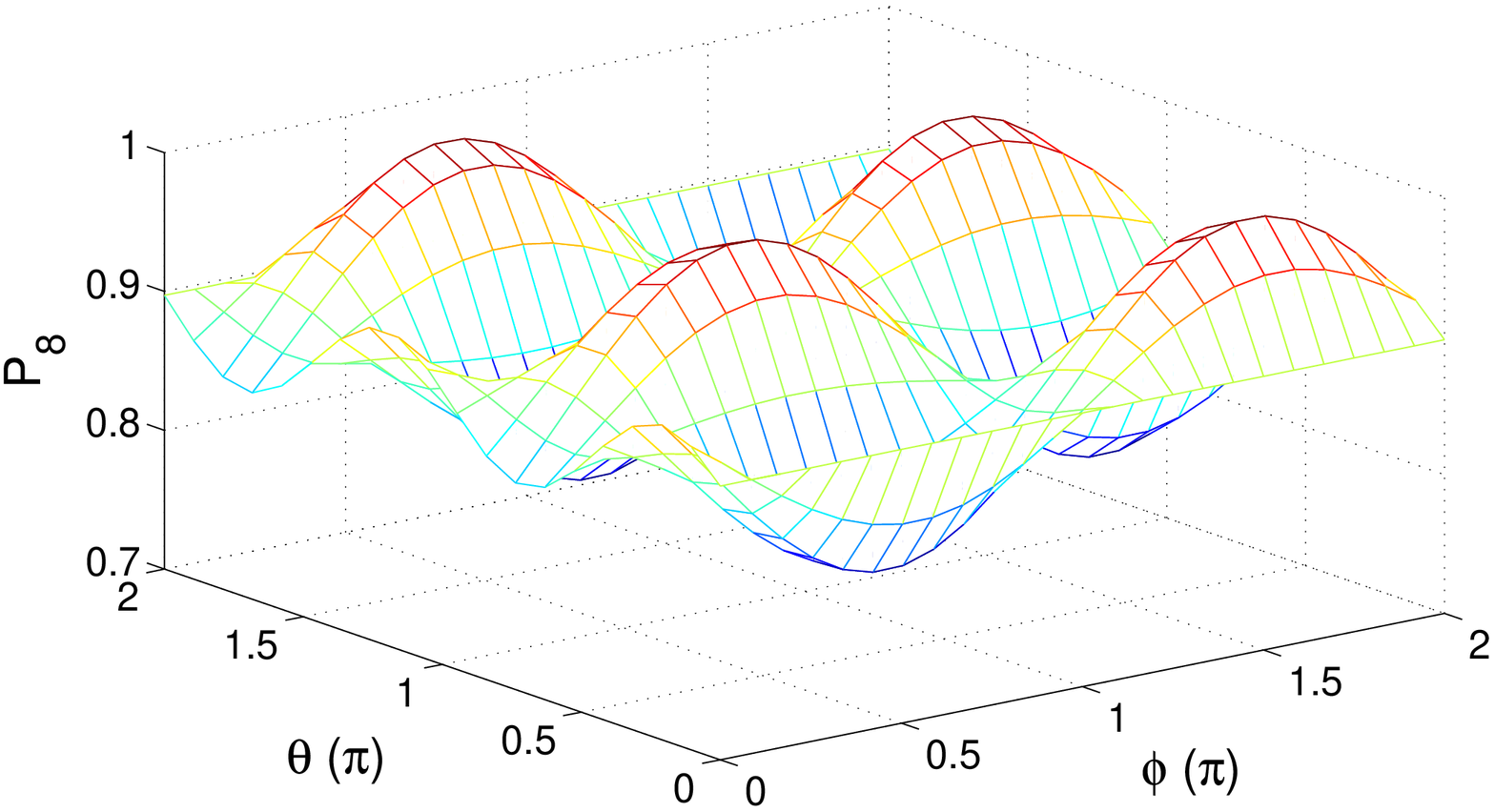}\caption{The dependence of the
transfer efficiency on the initial states. Top panel:  the initial
state site  is a superposition of $|1\rangle$  and $|6\rangle$.
Namely, $|\psi(t=0)\rangle=\cos\theta|1\rangle+\sin\theta\exp(i
\phi)|6\rangle.$ Lower panel: the initial state is
$|\psi(t=0)\rangle=\cos\theta|1\rangle+\sin\theta\exp(i
\phi)|2\rangle.$ The phases in $J_{ij}$  are optimized for maximal
transfer efficiency, i.e., they take the same values  as in
Fig.\ref{fig1}. } \label{fig2}
\end{figure}

\begin{figure}
\includegraphics*[width=0.8\columnwidth,
height=0.5\columnwidth]{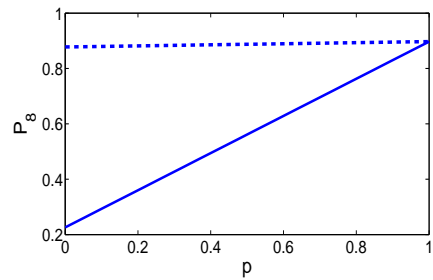} \caption{The transfer
efficiency versus initial states: the case of classical
superposition. The phases (couplings)  are the same as in Fig.
\ref{fig1}. The initial state is  $p|1\rangle\langle
1|+(1-p)|6\rangle\langle 6|$ for solid line, and it is $p
|1\rangle\langle 1|+(1-p)|2\rangle\langle 2|$ for dotted line.}
\label{fig3}
\end{figure}

Spatial and temporal relaxation of exciton  shows that the site 1
and 6 were populated initially with large contribution
\cite{adolphs06}. Then it is interesting to study how the initial
states affect the excitation transfer efficiency. In the following,
we shall shed light on this issue. Two cases are considered. First,
we calculate the excitation transfer efficiency with exciton
initially in a superposition of site 1 and 6 and analyze  the effect
of transfer efficiency on the initial states. Second, we extend this
study to initial states where sites 1 and 2 are initially excited.
These analyses are performed by numerically calculate the  transfer
efficiency at time $T= 5 $ps, selected results are shown in Fig.
\ref{fig2} and Fig. \ref{fig3}.

In Fig. \ref{fig2} we present the transfer efficiency as a function
of $\theta$ and $\phi$, which characterize the pure initial states
of the FMO complex through
$|\psi(t=0)\rangle=\cos\theta|1\rangle+\sin\theta\exp(i
\phi)|6\rangle$ (upper panel) and
$|\psi(t=0)\rangle=\cos\theta|1\rangle+\sin\theta\exp(i
\phi)|2\rangle$ (lower panel). We find the FMO complex efficient to
transfer excitation initially at site 1 is not good at EET with site
6 occupied. Namely, coherent superposition of sites 1 and 6
decreases the exciton transfer efficiency. For exciton initially in
a superposition of 1 and 6, the transfer efficiency is more
sensitive to the population ratio (characterized by $\theta$) but
not to the relative phase $\phi$.   For exciton initially excited on
site 1 and 2, both the population ratio and the relative phase
affect the energy transfer, the transfer efficiency reaches its
maximum with $\theta=\frac 3 4\pi$ and $\phi=\frac 1 2\pi$ and it
arrives at its minimum with $\theta=\frac {\pi}{4}$ and $\phi=\frac
1 2\pi$. This observation suggests that a properly superposition of
site 1 and site 2 can enhance  the EET, although the increase of
efficiency is small.

Fig. \ref{fig3} shows the dependence of the transfer efficiency on
the  mixing rate $p$, where the initial state is a classical mixing
of site 1 and 6 (or 2). Obviously, the population mixing of site 1
and site 6 does not favors the transfer efficiency. While  the
classical mixing of site 1 and site  2  increase the transfer
efficiency a little bit (not clear from the figure).

It is recently suggested that decoherence is  an essential feature
of EET in the FMO complex, where decoherence arises from
interactions with the protein cage, the reaction center and the
surrounding environment. Further examination  shown that dephasing
can increase the efficiency of transport in the FMO, whereas  other
types of decoherence (for example dissipation) block the EET. These
effects of decoherence  can be understood as  the
fluctuation-induced-broadening of energy levels, which bridges the
on-site energy gap and  changes equivalently  the coupling between
the sites.

We now examine if the decoherence can further increase the EET
efficiency with the complex inter-site couplings.  The decoherence
can be described by an added Liouvillian term
\begin{equation}
\mathcal{L}(\rho)=\sum_{j=1}^7 \gamma_j \left[P_j \rho(t)
P_j-\frac{1}{2}P_j\rho(t)-\frac{1}{2}\rho(t)P_j\right],
\label{masterE1}
\end{equation}
to Eq. (\ref{masterE}) with $P_j=|j\rangle\langle j|$,
$j=1,2,3,...,7$. The decoherence may come from site-environment
couplings, where the environment models the thermal phonon bath and
radiation fields.
\begin{figure}
\includegraphics*[width=0.8\columnwidth,
height=0.5\columnwidth]{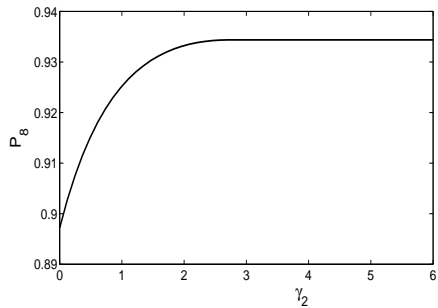} \caption{EET efficiency as a
function of $\gamma_2$. The other decoherence rates $\gamma_j$  are
zero, the phases in the inter-site couplings are optimized for EET
efficiency with $\gamma_2=0.$} \label{fig4}
\end{figure}
Optimizing  the decoherence rates  $\gamma_j, (j=1,2,...,7)$ with
$\Gamma=44$ for the EET  efficiency, we find   $\gamma_2=2.4$ and
$\gamma_j=0, (j\neq 2)$ yields a transfer efficiency $P_8=0.9344,$
see Fig. \ref{fig4}.  Non-zero $\gamma_j, (j=1,3,4,5,6,7)$ can
 increase the EET efficiency, but the amplitude of the
increase is not evident, for example, $\gamma_1=2.4, \gamma_3=2.0,
\gamma_6=1.8,$ the other $\gamma_j=0$ and $\Gamma=44$ yields
$P_8=0.9374$.

\section{effect of fluctuations}
The site energies are among the most debated properties of the FMO
complex. These values are needed for exciton calculations of the
linear spectra and simulations of dynamics. They depend on local
interactions between the {\it BChl a} molecule and the protein
envelope and include electrostatic interactions and ligation. Since
the interactions are difficult to identify and even harder to
quantify, the site energies are usually treated as independent
parameters  obtained from a simultaneous fit to several optical
spectra. There are many approaches to obtain the site energies by
fitting  the spectra.  One of the main differences between those
approaches is whether they restrict the interactions to {\it BChl a}
molecules within a subunits or whether they include interactions in
the whole trimer. Since the site energies and inter-site couplings
may be different and fluctuate due to the couplings to its
surroundings, it is interesting to ask how the fluctuations in the
site energies and couplings affect the transfer efficiency. In the
following, we will study this issue and show that the exciton
transfer in the FMO complex remains essentially unaffected in the
presence of random variations in site
 energies and inter-site couplings.  This strongly suggests that the
experimental results recorded for samples at low temperature would
also be observable at higher temperatures.

To be specific, we consider two types of fluctuations in the site
energies and inter-site couplings. In the first one, it has zero
mean, while the another type of fluctuations has nonzero positive
mean. For the fluctuations with zero mean, the Hamiltonian in Eq.
(\ref{ha}) takes the following changes, $H_{jj}\rightarrow
H_{jj}(1+r_1\cdot(\mbox {rand}(1)-0.5))$ and $H_{i\neq j}\rightarrow
H_{i\neq j}(e^{i\phi_{ij}}+r_2\cdot(\mbox {rand}(1)-0.5)).$ For the
fluctuations with nonzero positive mean, the Hamiltonian takes the
following changes, $H_{jj}\rightarrow H_{jj}(1+r_1\cdot \mbox
{rand}(1))$  and $H_{i\neq j}\rightarrow H_{i\neq
j}(e^{i\phi_{ij}}+r_2\cdot\mbox {rand}(1)),$ where $\mbox{rand(1)}$
denotes a random number between 0 and 1. So a 100\% static disorder
may appear in the on-site energies and inter-site couplings.

\begin{figure}
\includegraphics*[width=0.8\columnwidth,
height=0.5\columnwidth]{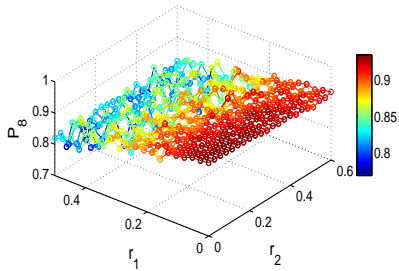} \caption{Effect of fluctuations
in the Hamiltonian on the transfer efficiency. The plot is for  the
fluctuations with zero mean. The resulting $P_8$ is an averaged
result over 20 independent runs. The fluctuations with non-zero mean
has a similar effect on the EET, which is not shown here.}
\label{fig5}
\end{figure}

With these arrangements, we numerically calculated the transfer
efficiency and present the results in Fig. \ref{fig5}. Each value of
transfer efficiency is a result averaged over 20 fluctuations. From
the numerical simulations, we notice that the results for
fluctuations with non-zero mean are almost the same as those for
zero-mean fluctuations, we hence present here only the results for
zero-mean fluctuations.  Two observations are obvious. (1) With
zero-mean fluctuations in the site energies and couplings increase,
the transfer efficiency fluctuates greatly. (2) The efficiency
decreases with both $r_2$ and $r_1$. Moreover, the EET efficiency is
sensitive to the fluctuations in the site energy more than that in
the inter-site couplings.

These feature  can be interpreted as follows.  Since  the network
with Hamiltonian Eq.(\ref{ha}) is optimal for EET efficiency, any
changes in the site-energy and inter-site couplings would diminish
the excitation energy transfer, leading to the decrease in the EET
efficiency. This feature is different from the case  that the
decoherence dominates the mechanism of EET. As Ref. \cite{yi12}
shown, the efficiency increases with $r_1$ but decreases with $r_2$.
The physics behind this difference is as follows.  For the case that
the decoherence dominates the mechanism of EET, the energy gap
between neighboring sites blocks the energy transfer, whereas the
inter-site couplings (representing the overlap of different sites)
together with the decoherence  favor the transport. Whereas for the
present case  the phases in the inter-site couplings need to {\it
match} the energy spacing between sites. As a result, any mismatches
would results in decrease in the EET efficiency.

\begin{figure}
\includegraphics*[width=0.8\columnwidth,
height=0.5\columnwidth]{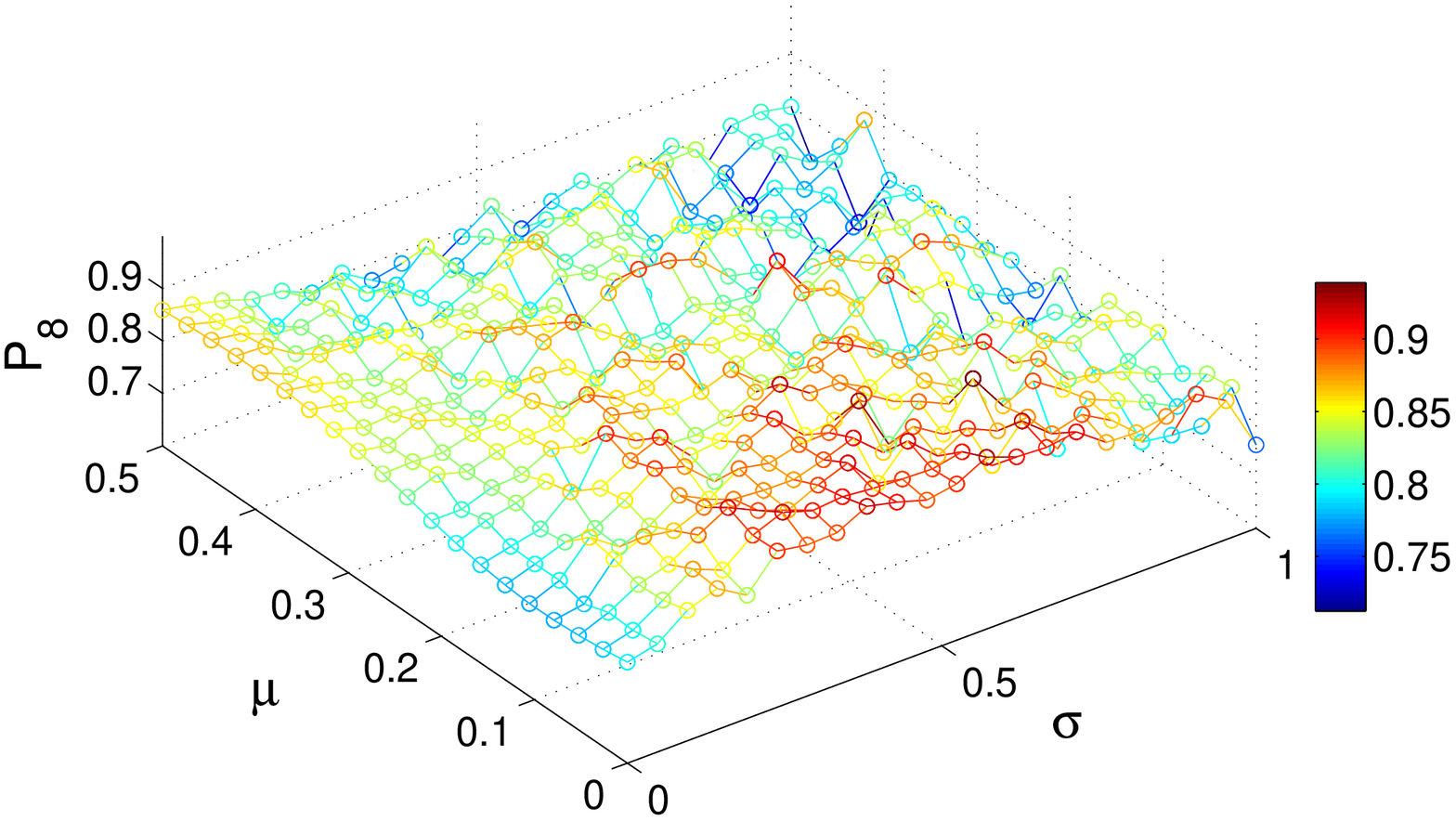}
\includegraphics*[width=0.8\columnwidth,
height=0.5\columnwidth]{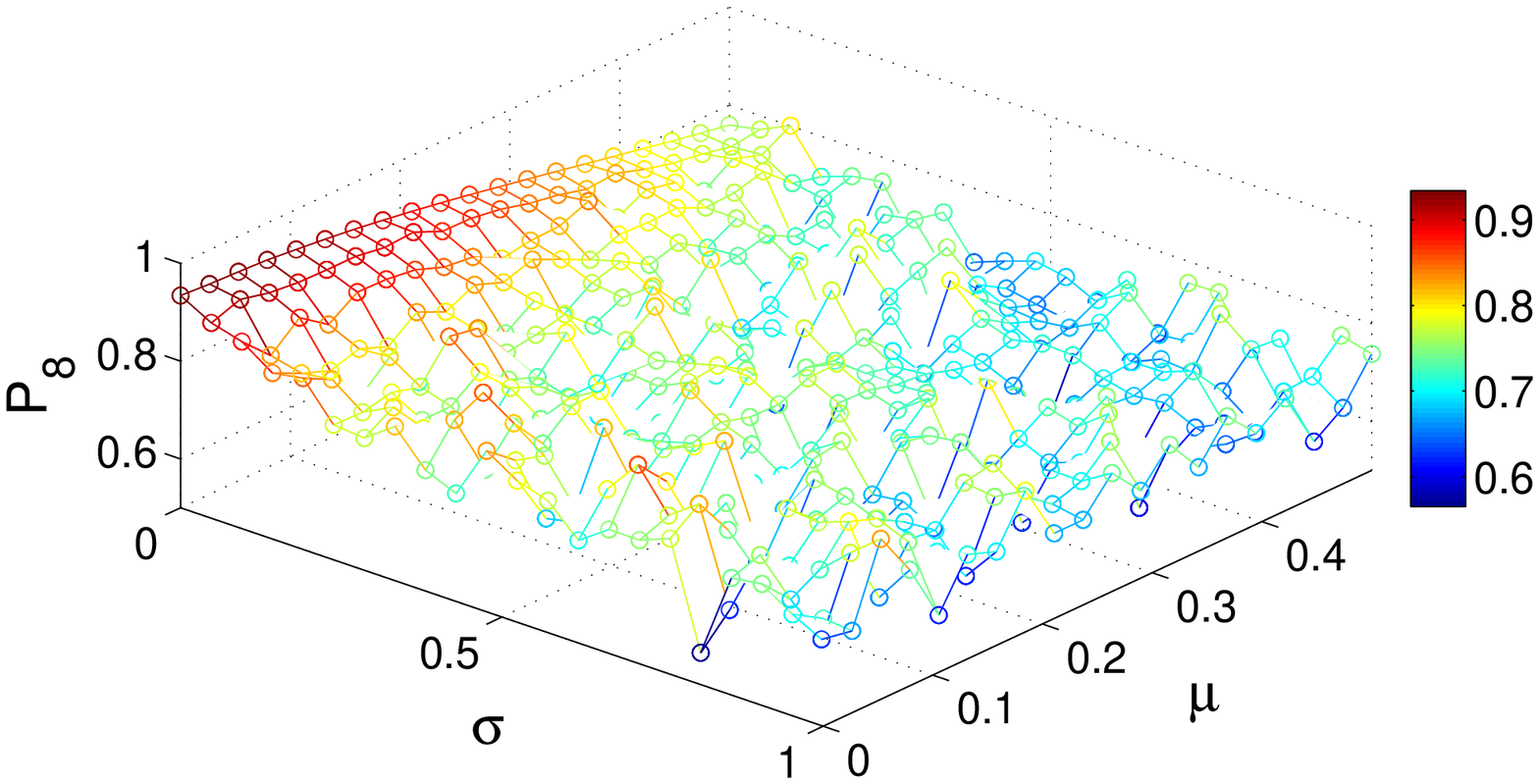} \caption{Effect of fluctuations
in the Hamiltonian on the transfer efficiency.  The fluctuations are
Gaussian with mean $\mu$  and the standard deviation $\sigma$.
Upper: The fluctuations happen only in the on-site energies; Bottom:
The fluctuations in both the on-site energies and couplings. The
resulting fidelity is an averaged result over 15 independent runs.}
\label{fig6}
\end{figure}

It was suggested\cite{adolphs06}  that the fluctuations in the
on-site energies are Gaussian. To examine the effect of Gaussian
fluctuations, we introduce a Gaussian function,
$y(x|\sigma,\mu)=\frac{1}{\sigma\sqrt{2\pi}}e^{-\frac{(x-\mu)^2}{2\sigma^2}},$
where $\mu$ is the mean, while $\sigma$ denotes the standard
deviation. We assume that the fluctuations enter only into  the
on-site energies, namely,  $H_{jj}$ is replaced by
$H_{jj}(1+y(x|\sigma,\mu))$, $j=1,2,...,7$. With these notations, we
plot the effect of fluctuations on the transfer efficiency in Fig.
\ref{fig6}(upper panel). We find that the dependence of the
efficiency on  the variation is subtle: When the mean is zero, the
efficiency reach its maximum at $\sigma=0.5,$ while when the mean is
large, small variation favors the EET efficiency. Moreover, we find
that the efficiency  decreases as the mean increases. This again can
be understood as  mismatch  between the site energy and the
inter-site couplings caused by the fluctuations. When these Gaussian
fluctuations occur in both the on-site energies and the couplings,
we find from Fig.\ref{fig6}(bottom panel) that the transfer
efficiency decreases as the mean  and variation increases.

Absorption spectroscopy and 2D echo-spectroscopy are used to study
time-resolved processes such as energy transfer or vibrational decay
as well as to measure intermolecular couplings strengths. They
provide a tool that gives direct insight into the energy states of a
quantum system. In turn,  the eigenenergies are essential  to
calculate the absorption spectroscopy and 2D echo-spectroscopy. It
is easy to find that the eigenenergies and energy gaps for the FMO
with complex inter-site couplings, given by (96.2851, 62.2958,
61.4307, 48.2046, 22.6379, 18.9438,  -4.1374),  are almost the same
as in the system with real inter-site couplings given by (96.8523,
62.6424, 57.9484, 50.6357, 22.8218, 19.2387,  -4.4789). The maximal
difference is at the 7th eigenenergy, it is about 8.25\% to its
eigenenergy. This suggests that the absorption spectroscopy and 2D
echo-spectroscopy may not change much due to the introduced phases
in the inter-site couplings.

\section{Conclusion}
Excitation  energy transfer (EET) has been an interesting subject
for decades not only for its phenomenal efficiency but also for its
fundamental role in Nature.  Recent studies show that the quantum
coherence itself cannot explain the very high excitation energy
transfer efficiency in the Fenna-Matthews-Olson  Complex. In this
paper, we show that this is not the case when the inter-site
couplings are complex. Based on the Frenkel exciton Hamiltonian and
the phenomenologically introduced decoherence, we have optimized the
phases in the inter-site couplings  for maximal energy transfer
efficiency, which can reach about 89.72\% at time 5ps without
decoherence and 93.44\% with only dephasing at site 2. By
considering different mixing of exciton on site 1 and site 6 (site
2) as the initial states, we have examined the effect of initial
states on the energy transfer efficiency. The results suggest that
any mixing of site 1 and site 6 decreases the energy transfer
efficiency, whereas a coherent superposition (or classically mixing)
of site 1 and site 2 does not change much the EET efficiency. The
fluctuations in the site energies and inter-site couplings diminish
the EET due to the mismatch caused by the fluctuations.

Finishing this work, we have noticed a closely related
preprint\cite{zhu12}. While they focus on bi-pathway EET in the FMO
(i.e., neglecting the inter-site couplings weaker than 15 $cm^{-1}$)
and a general representation for complex network, we are mainly
concerned about the phases added to the inter-site couplings,
initial excitations and fluctuations in the site energies and
inter-site couplings. In this respect, both works are complementary
to each other.

\ \ \\
Stimulating discussions with Jiangbin Gong are acknowledged. This
work is supported by the NSF of China under Grants Nos 61078011 and
11175032 as well as the National Research Foundation and Ministry of
Education, Singapore under academic research grant No. WBS:
R-710-000-008-271.


\begin{references}
\bibitem{olson63} C. Sybesma and J. M. Olson, Proc. Natl Acad. Sci.
USA {\bf 49}, 248(1963).

\bibitem{fenna75} R. E. Fenna and B. W. Mattews, Nature {\bf 258},
573 (1975).

\bibitem{engel07} G. S. Engel, T.R. Calhoun, E.L. Read, T.-K. Ahn, T. Manal, Y.-C.
Cheng, R.E. Blankenship, and G.R. Fleming, Nature 446, 782 (2007).

\bibitem{lee07} H. Lee, Y. C. Cheng, and G. R. Fleming, Science 316, 1462
(2007); V. I. Prokhorenko et al., J. Phys. Chem. B 106, 9923 (2002).

\bibitem{adolphs06} J. Adolphs and T. Renger, Biophys. J. 91, 2778, (2006).

\bibitem{mohseni08} M. Mohseni, P. Rebentrost, S. Lloyd, and A. Aspuru-Guzik, J.
Chem. Phys. 129, 174106 (2008).

\bibitem{plenio08} M. B. Plenio and S.F. Huelga, New J. Phys. 10, 113019 (2008).

\bibitem{caruso09} F. Caruso, A.W. Chin, A. Datta, S.F. Huelga, and M.B. Plenio, J.
Chem. Phys. 131, 105106 (2009).

\bibitem{chin10} A. W. Chin, A. Datta, F. Caruso, S.F. Huelga, and M.B. Plenio, New
J. Phys. 12, 065002 (2010).

\bibitem{olaya08} A. Olaya-Castro, C.F. Lee, F.F. Olsen, and N.F. Johnson, Phys.
Rev. B 78, 085115 (2008).

\bibitem{ishizaki09} A. Ishizaki and G.R. Fleming, Proc. Natl. Acad. Sci. 106, 17255
(2009).

\bibitem{yang10}  S. Yang, D. Z. Xu, Z. Song, C. P. Sun, J. Chem. Phys. 132, 234501
(2010).

\bibitem{hoyer09} S. Hoyer, M. Sarovar, and K.B. Whaley, e-print arXiv:0910.1847
(2009).

\bibitem{sarovar11} M. Sarovar, Y. C. Cheng, and K. B. Whaley, Phys.
Rev. E {\bf 83}, 011906 (2011).

\bibitem{shi11} R. Zheng, Y. Jing, and Q. Shi, J. Chem. Phys. {\bf
134}, 194508(2011).

\bibitem{caruso101} F. Caruso, S.F. Huelga, and M.B. Plenio, e-print arXiv:1003.5877
(2010).

\bibitem{shabani11} A. Shabani,  M. Mohseni, H. Rabitz,
and S. Lloyd, e-print arXiv: 1103.3823.

\bibitem{ghosh11} P. K. Ghosh, A.Yu. Smirnov, F. Nori,  J. Chem.
Phys. 134, 244103 (2011).

\bibitem{yi11} B. Cui, X. Y. Zhang, X. X. Yi, e-print arXiv:1106.4429.

\bibitem{adolphs08} J. Adolphs, F.  M\"uh, M. E. Madjet, T. Renger,
Photosynth Res {\bf 95}, 197 (2008).

\bibitem{tronrud09} D. E. Tronrud, J. Wen, L. Gay and R. E.
Blankenship, Photosynth. Res. {\bf 100}, 79 (2009).

\bibitem{busch11} M. S. Busch, F. Muh, M. E. Madjet, and T. Renger,
J. Phys. Chem. Lett. {\bf 2}, 93 (2011).

\bibitem{ritschel11}  G. Ritschel, J. Roden, W. T. Strunz, A.
A. Guzik, and A. Eisfeld, J. Phys. Chem. Lett. {\bf 2}, 2912 (2011).

\bibitem{yi12} B. Cui, X. X. Yi, and C. H. Oh, J. Phys. B: At. Mol. Opt. Phys. {\bf 45},
085501(2012).

\bibitem{zhu12} Bao-quan Ai and Shi-Liang Zhu, e-print arXiv: 1208.2778.
\end{references}
\end{document}